# The Evaluation of Causal Effects in Studies with an Unobserved Exposure/Outcome Variable: Bounds and Identification

**Manabu Kuroki**
Department of Systems Innovation
Graduate School of Engineering Science
Osaka University
mkuroki@sigmath.es.osaka-u.ac.jp

**Zhihong Cai**
Department of Biostatistics
School of Public Health
Kyoto University
cai@pbh.med.kyoto-u.ac.jp

## Abstract

This paper deals with the problem of evaluating the causal effect using observational data in the presence of an unobserved exposure/outcome variable, when cause-effect relationships between variables can be described as a directed acyclic graph and the corresponding recursive factorization of a joint distribution. First, we propose identifiability criteria for causal effects when an unobserved exposure/outcome variable is considered to contain more than two categories. Next, when unmeasured variables exist between an unobserved outcome variable and its proxy variables, we provide the tightest bounds based on the potential outcome approach. The results of this paper are helpful to evaluate causal effects in the case where it is difficult or expensive to observe an exposure/outcome variable in many practical fields.

## 1 INTRODUCTION

The evaluation of causal effects from observational studies is one of the central aims in many fields of practical science. For this purpose, many researchers have attempted to clarify cause-effect relationships and to evaluate the causal effect of an exposure variable on an outcome variable through observed data. Statistical causal analysis, which is one of powerful tools for solving these problems, started with path analysis (Wright, 1923, 1934), and advanced to structural equation models (Wold, 1954; Bollen, 1989). It also has been modified in order to be applicable to categorical data (Goodman, 1973, 1974a, 1974b; Hagenaars, 1993). Recently, Pearl (2000) developed a new framework of causal modeling based on a directed acyclic graph and the corresponding nonparametric structural equation model.

In observational studies, there often exist unobserved variables, which makes it difficult to evaluate reliable causal effects. Many researchers have proposed various useful approaches to evaluate causal effects when unobserved variables are confounding factors between an exposure variable and an outcome variable, such as the instrumental variable method and sensitivity analysis. In the context of graphical causal models, Pearl (2000) provided the mathematical definition of the causal effect. In addition, when both an exposure variable and an outcome variable are observed, Pearl (2000), Tian and Pearl (2002) and Shpitser and Pearl (2006) discussed several graphical identification conditions for causal effects, which enable us to recognize situations where the causal effects can be evaluated from observational data.

However, in some situations, even an exposure/outcome variable is unobserved. For example, in a study to examine whether the socioeconomic gradient has an influence on low birth-weight, socioeconomic status is measured by some proxy variables such as income, wealth, education and occupation, since the true socioeconomic status is unobserved (Finch, 2003). Another example concerning an unobserved exposure is in occupational settings. Many epidemiological studies have addressed the question of carcinogenicity in workers exposed to diesel exhaust and coal mine dust, and most showed a low-to-medium increase in the risk of lung cancer. However, exposure measurement in these studies is mainly inferred on the basis of job classifications and may lead to misclassification (Hoffmann and Jockel, 2006). On the other hand, as an example concerning an unobserved outcome, Fleiss et al. (1976) reported a comparative clinical trial of ibuprofen, aspirin and placebo in the relief of postextraction pain. Since the true outcome (pain relief) is unobserved, they used the Ridit analysis (Bross, 1958) to divide patients into five categories of pain relief: none, poor, fair, good and very good. These examples show the importance of evaluating causal effects when an exposure/outcome variable is unobserved.

Kuroki et al. (2005) pointed out that it is difficult to apply the identification criteria proposed by Pearl and his colleagues to evaluate causal effects in such situations, and provided the graphical identifiability criteria when an unobserved exposure/outcome variable is continuous. In addition, Kuroki (2007) arranged the identification conditions proposed by Kuroki et al. (2005) to the case where an exposure/outcome variable is dichotomous. However, in many situations, researchers and practitioners are more interested in the different exposure levels (e.g., none, low, medium and high) than the pure binary exposure (exposed vs. unexposed), and are also more interested in the response levels (e.g., none, poor, fair, good and very good) than the simple binary response (improved vs. not improved).

Then, the main purpose of this paper is to provide identifiability criteria for causal effects from observational studies in the presence of an unobserved exposure/outcome variable with more than two categories. It will be shown that if we can observe some proxy variables that are affected by the unobserved variable, then the causal effect can be evaluated by using statistical causal analysis. More generally, we consider the case where there exist unmeasured variables between the unobserved exposure/outcome variable and its proxy variables. Under such a situation, the causal effect is not identifiable but the bounds on the causal effect can be derived. Finally, we illustrate our results with an example about social science.

## 2  PRELIMINARIES

### 2.1  BAYESIAN NETWORKS

Let $f(v_1, v_2, \ldots, v_n)$ be a strictly positive joint distribution of a set $\boldsymbol{V} = \{V_1, V_2, \cdots, V_n\}$ of variables, $f(v_i|v_j)$ the conditional distribution of $V_i$ given $V_j = v_j$ $(V_i, V_j \in \boldsymbol{V})$ and $f(v_i)$ the marginal distribution of $V_i$. Similar notations are used for other distributions. For graph theoretic terminology used in this paper, refer to Kuroki et al. (2005).

Suppose that a set $\boldsymbol{V}$ of variables and a directed acyclic graph $G = (\boldsymbol{V}, \boldsymbol{E})$ are given. When the joint distribution of $\boldsymbol{V}$ is factorized recursively according to the graph $G$ as the following equation, the graph is called a Bayesian network:

$$f(v_1, v_2, \cdots, v_n) = \prod_{i=1}^{n} f(v_i|\text{pa}(v_i)). \qquad (1)$$

When $\text{pa}(v_i)$ is an empty set, $f(v_i|\text{pa}(v_i))$ is the marginal distribution $f(v_i)$ of $v_i$.

If a joint distribution is factorized recursively according to the graph $G$, the conditional independencies implied by the factorization (1) can be obtained from the graph $G$ according to the d-separation criterion (Pearl, 1988), that is, if $\boldsymbol{Z}_1$ d-separates $\boldsymbol{Z}_2$ from $\boldsymbol{Z}_3$ in a directed acyclic graph $G$ $(\boldsymbol{Z}_1, \boldsymbol{Z}_2, \boldsymbol{Z}_3 \subset \boldsymbol{V})$, then $\boldsymbol{Z}_2$ is conditionally independent of $\boldsymbol{Z}_3$ given $\boldsymbol{Z}_1$ in the corresponding recursive factorization (1); See, for example, Geiger et al. (1990).

### 2.2  CAUSAL EFFECT

Pearl (2000) defined a causal effect as a distribution of an outcome variable when conducting an external intervention, where an 'external intervention' means that a variable is forced to take on some fixed value, regardless of the values of other variables. If the distribution of the remaining variables represented in the directed acyclic graph remains essentially unchanged by such an external intervention, then the graph can be regarded as a causal diagram and the effect of the external intervention can be calculated from the joint factorized distribution. The exact definition is given as follows.

**DEFINITION 1**
Let $\boldsymbol{V} = \{X, Y\} \cup \boldsymbol{Q}$ $(\{X, Y\} \cap \boldsymbol{Q} = \phi)$ be a set of variables represented in a Bayesian network $G$. If the distribution of $Y$ after setting $X$ to a value $x$ is given by

$$f(y|\text{set}(X = x)) = \sum_{\boldsymbol{q}} \frac{f(x, y, \boldsymbol{q})}{f(x|\text{pa}(x))}, \qquad (2)$$

then $G$ is called a causal diagram with regard to $X$ and equation (2) is called a causal effect of $X$ on $Y$. Here, $\text{set}(X = x)$ means that $X$ is set to a value $x$ by an external intervention. □

If Definition 1 holds true with regard to all pairs of variables in the graph, then the whole graph is said to be causal. For more details about the relationship between Bayesian networks and causal diagrams, see Pearl (2000).

Given a causal diagram $G$, in order to evaluate the causal effect $f(y|\text{set}(X = x))$ of $X$ on $Y$ from a joint factorized distribution of observed variables, it is required to observe not only $X$ and $Y$ but also a set $\boldsymbol{Z}$ of other variables, such as confounders. Pearl (2000) provided 'the back door criterion' as one of graphical identifiability criteria for causal effects $f(y|\text{set}(X = x))$, where 'identifiable' means that $f(y|\text{set}(X = x))$ can be determined uniquely from a joint distribution of observed variables.

**DEFINITION 2**
Suppose that $X$ is a non-descendant of $Y$ in a directed acyclic graph $G$. If a set $\boldsymbol{Z}$ of vertices satisfies the following conditions relative to an ordered pair $(X, Y)$ of vertices, then $Z$ is said to satisfy the back door criterion relative to $(X, Y)$:

(i) no vertex in $Z$ is a descendant of $X$;

(ii) $Z$ blocks every path between $X$ and $Y$ that contains an arrow pointing to $X$. □

If a set $Z$ of variables satisfies the back door criterion relative to $(X, Y)$, then the causal effect $f(y|\text{set}(X = x))$ of $X$ on $Y$ is identifiable through the observation of $Z \cup \{X, Y\}$ and is given by the formula

$$f(y|\text{set}(X = x)) = \sum_{z} f(y|x, z) f(z). \quad (3)$$

When the back door criterion can not be applied to evaluate causal effects, Pearl (2000) provided 'the front door criterion', which is as follows:

**DEFINITION 3**
Suppose that $X$ is a non-descendant of $Y$ in a directed acyclic graph $G$. If a set $Z$ of variables satisfies the following conditions relative to an ordered pair $(X, Y)$ of variables, then $Z$ is said to satisfy the front door criterion relative to $(X, Y)$:

(i) $Z$ blocks all directed paths from $X$ to $Y$;

(ii) an empty set blocks every path between $X$ and $Z$ that contains an arrow pointing to $X$;

(iii) $X$ blocks every path between any vertex in $Z$ and $Y$. □

If a set $Z$ of variables satisfies the front door criterion relative to $(X, Y)$, then the causal effect $f(y|\text{set}(X = x))$ of $X$ on $Y$ is identifiable through the observation of $Z \cup \{X, Y\}$ and is given by the formula

$$f(y|\text{set}(X = x)) = \sum_{x', z} f(y|x', z) f(z|x) f(x'). \quad (4)$$

## 3 IDENTIFICATION OF CAUSAL EFFECTS

In section 2, it is assumed that both an exposure variable and an outcome variable are observable. If either of them is unobserved, we cannot identify the causal effect of an exposure on an outcome even if a set of variables satisfying the back door criterion or the front door criterion are observed. In this section, we consider the case where an unobserved exposure/outcome variable is assumed to be discrete. Let $X$ be an exposure variable and $Y$ be an outcome variable. Though $X$ or $Y$ is unobserved, researchers are interested in dividing them into $k$ categories. For example, when the domain of $Y$ is divided into $k = 3$ categories, $y_1$, $y_2$ and $y_3$ may represent the poor, fair and good response levels. Then, let $U$ be either $X$ or $Y$ which is an unobserved variable ($u \in \{u_1, \cdots, u_k\}$). In addition, let a set $S$ and a set $T$ be observed proxy variables that are affected by the unobserved variable $U$. Assume that we can select $k$ distinct vectors from the domains of a set $S$ and a set $T$ of variables, denoted as $t_1, \cdots, t_k$ and $s_1, \cdots, s_k$, respectively. A set $W$ and a set $Z$ are assumed to be continuous and/or discrete variables. Furthermore, let $P$ and $Q$ be $k$ dimensional nonsingular matrices such that

$$P = \begin{pmatrix} 1 & f(t_1|z) & \cdots & f(t_{k-1}|z) \\ f(s_1|z) & f(s_1, t_1|z) & \cdots & f(s_1, t_{k-1}|z) \\ \vdots & \vdots & \vdots & \vdots \\ f(s_{k-1}|z) & f(s_{k-1}, t_1|z) & \cdots & f(s_{k-1}, t_{k-1}|z) \end{pmatrix},$$
(5)

$$Q = \begin{pmatrix} f(w|z) & f(w, t_1|z) & \cdots & f(w, t_{k-1}|z) \\ f(w, s_1|z) & f(w, s_1, t_1|z) & \cdots & f(w, s_1, t_{k-1}|z) \\ \vdots & \vdots & \vdots & \vdots \\ f(w, s_{k-1}|z) & f(w, s_{k-1}, t_1|z) & \cdots & f(w, s_{k-1}, t_{k-1}|z) \end{pmatrix}.$$
(6)

Then, the following theorem is obtained.

**THEOREM 1**
Given a causal diagram $G$ on $V$ with $S \cup T \cup \{U\} \cup Z \cup W (\subset V)$, suppose that

(i) $Z \cup \{U\}$ d-separates $S$ from $T$ and $W$ from $S \cup T$;

(ii) $f(u_1|z) < \cdots < f(u_k|z)$ holds true for any $z$;

(iii) For the matrices defined as equations (5) and (6), both $P$ and $Q$ are $k$ dimensional nonsingular matrices and $|Q - \lambda P| = 0$ has non-zero distinct solutions of $\lambda$ ($0 < \lambda_1 < ... < \lambda_k$) for any $z$ ($P \neq Q$),

then the distribution $f(u, w, z)$ is identifiable through the observation of $S \cup T \cup Z \cup W$. □

**PROOF OF THEOREM 1**
Let

$$P_1 = \begin{pmatrix} 1 & f(t_1|u_1, z) & \cdots & f(t_{k-1}|u_1, z) \\ \vdots & \vdots & \vdots & \vdots \\ 1 & f(t_1|u_k, z) & \cdots & f(t_{k-1}|u_k, z) \end{pmatrix},$$

$$P_2 = \begin{pmatrix} 1 & f(s_1|u_1, z) & \cdots & f(s_{k-1}|u_1, z) \\ \vdots & \vdots & \vdots & \vdots \\ 1 & f(s_1|u_k, z) & \cdots & f(s_{k-1}|u_k, z) \end{pmatrix},$$

and $\Delta = \text{diag}(f(w|u_1, z), \cdots, f(w|u_k, z))$ be the $k$ dimensional diagonal matrices of conditional probabili-ties of observed variables $\boldsymbol{W}$ given $U$ and $\boldsymbol{Z}$. In addition, let $M = \mathrm{diag}(f(u_1|\boldsymbol{z}), \cdots, f(u_k|\boldsymbol{z}))$ be the $k$ dimensional diagonal matrix of conditional probabilities of $U$ given $\boldsymbol{Z}$. Then, the followings are derived:

$$f(\boldsymbol{w}|\boldsymbol{z}) = \sum_{i=1}^{k} f(\boldsymbol{w}|u_i, \boldsymbol{z}) f(u_i|\boldsymbol{z}),$$

$$f(\boldsymbol{t}_j, \boldsymbol{s}_l|\boldsymbol{z}) = \sum_{i=1}^{k} f(\boldsymbol{s}_l|u_i, \boldsymbol{z}) f(\boldsymbol{t}_j|u_i, \boldsymbol{z}) f(u_i|\boldsymbol{z}),$$

$$f(\boldsymbol{w}, \boldsymbol{t}_j, \boldsymbol{s}_l|\boldsymbol{z}) = \sum_{i=1}^{k} f(\boldsymbol{w}|u_i, \boldsymbol{z}) f(\boldsymbol{t}_j|u_i, \boldsymbol{z})$$
$$\times f(\boldsymbol{s}_l|u_i, \boldsymbol{z}) f(u_i|\boldsymbol{z}) \quad (7)$$

for $j, l = 1, 2, \cdots, k$. Then, we can obtain

$$P = P_2' M P_1 \quad \text{and} \quad Q = P_2' M \Delta P_1.$$

Thus, by noting that both $P$ and $Q$ are nonsingular matrices of conditional probabilities of observed variables, consider the following equation for $\lambda$:

$$|Q - \lambda P| = |P_1' M \Delta P_2 - \lambda P_1' M P_2|$$
$$= |P_1'||M||\Delta - \lambda I_k||P_2| = 0, \quad (8)$$

where $I_k$ is a $k$ dimensional identity matrix. By solving equation (8), we can obtain the element $f(\boldsymbol{w}|u_i, \boldsymbol{z})$ $(i = 1, 2, \cdots, k)$ of $\Delta$. Here, let $\lambda_i$ be a disjoint solution of the above equation satisfying $0 < \lambda_1 < \cdots < \lambda_k$. This means that $\Delta$ is identifiable if the order of $f(\boldsymbol{w}|u_1, \boldsymbol{z}), \cdots, f(\boldsymbol{w}|u_k, \boldsymbol{z})$ is known. Let $E_i = \mathrm{diag}(\alpha_{1(i)}, \cdots, \alpha_{k(i)})$ be a $k$ dimensional diagonal matrix $(i = 1, 2)$ and $A_1 = P_1^{-1} E_1$ be a $k$ dimensional matrix. Since $\Delta$, $E_1$ and $M$ are diagonal matrices and the elements of $\Delta$ are correspondent to the solutions of equation (8), we can obtain

$$Q A_1 = P A_1 \Delta.$$

This means that matrix $A_1 = P_1^{-1} E_1$ is the solution of the characteristic equation

$$(Q - \lambda P) \boldsymbol{x} = \boldsymbol{0}_k$$

for $\boldsymbol{x}$, which indicates that $A_1$ is estimable ($\lambda \in \{\lambda_1, \cdots, \lambda_k\}$). Here, $\boldsymbol{0}_k$ is a $k$ dimensional zero vector.

Since $P_1 = E_1 A_1^{-1}$, letting $A_1^{-1} = (a_1^{i,j})$, we can obtain

$$P_1 = \begin{pmatrix} 1 & f(\boldsymbol{t}_1|u_1, \boldsymbol{z}) & \cdots & f(\boldsymbol{t}_{k-1}|u_1, \boldsymbol{z}) \\ \vdots & \vdots & \vdots & \vdots \\ 1 & f(\boldsymbol{t}_1|u_k, \boldsymbol{z}) & \cdots & f(\boldsymbol{t}_{k-1}|u_k, \boldsymbol{z}) \end{pmatrix}$$
$$= E_1 A_1^{-1}$$
$$= \begin{pmatrix} \alpha_{1(1)} a_1^{1,1} & \alpha_{1(1)} a_1^{1,2} & \cdots & \alpha_{1(1)} a_1^{1,k} \\ \alpha_{2(1)} a_1^{2,1} & \alpha_{2(1)} a_1^{2,2} & \cdots & \alpha_{2(1)} a_1^{2,k} \\ \vdots & \vdots & \vdots & \vdots \\ \alpha_{k(1)} a_1^{k,1} & \alpha_{k(1)} a_1^{k,2} & \cdots & \alpha_{k(1)} a_1^{k,k} \end{pmatrix}.$$

Then, $\alpha_{i(1)} = 1/a_1^{i,1}$ can be uniquely obtained $(i = 1, 2, \cdots, k)$ from the first column, which indicates that $P_1$ is also estimable from $E_1 A^{-1}$ according to the order of $\lambda_1, \cdots, \lambda_k$, where $E_1 = \mathrm{diag}(1/a_1^{1,1}, \cdots, 1/a_1^{k,1})$.

On the other hand, letting $A_2 = P_2^{-1} E_2$, since $\Delta$, $E_2$ and $M$ are diagonal matrices and the elements of $\Delta$ are correspondent to the solutions of (8), we can obtain

$$Q' A_2 = P' A_2 \Delta.$$

This means that matrix $A_2 = P_2^{-1} E_2$ is the solution of the characteristic equation

$$(Q' - \lambda P') \boldsymbol{x} = \boldsymbol{0}_k$$

for $\boldsymbol{x}$. Thus, $A_2$ is also estimable ( $\lambda \in \{\lambda_1, \cdots, \lambda_k\}$).

Since $P_2 = E_2 A_2^{-1}$, letting $A_2^{-1} = (a_2^{i,j})$, we can obtain

$$P_2 = \begin{pmatrix} 1 & f(\boldsymbol{s}_1|u_1, \boldsymbol{z}) & \cdots & f(\boldsymbol{s}_{k-1}|u_1, \boldsymbol{z}) \\ \vdots & \vdots & \vdots & \vdots \\ 1 & f(\boldsymbol{s}_1|u_k, \boldsymbol{z}) & \cdots & f(\boldsymbol{s}_{k-1}|u_k, \boldsymbol{z}) \end{pmatrix}$$
$$= E_2 A_2^{-1}$$
$$= \begin{pmatrix} \alpha_{1(2)} a_2^{1,1} & \alpha_{1(2)} a_2^{1,2} & \cdots & \alpha_{1(2)} a_2^{1,k} \\ \alpha_{2(2)} a_2^{2,1} & \alpha_{2(2)} a_2^{2,2} & \cdots & \alpha_{2(2)} a_2^{2,k} \\ \vdots & \vdots & \vdots & \vdots \\ \alpha_{k(2)} a_2^{k,1} & \alpha_{k(2)} a_2^{k,2} & \cdots & \alpha_{k(2)} a_2^{k,k} \end{pmatrix}.$$

Then, $\alpha_{i(2)} = 1/a_2^{i,1}$ can be uniquely obtained $(i = 1, 2, \cdots, k)$ from the first column, which indicates that $P_2$ is also estimable from $E_2 A_2^{-1}$ according to the order of $\lambda_1, \cdots, \lambda_k$, where $E_2 = \mathrm{diag}(1/a_2^{1,1}, \cdots, 1/a_2^{k,1})$. From these results, we can obtain

$$P_2'^{-1} P P_1^{-1} = P_2'^{-1} (P_2' M P_1) P_1^{-1} = M. \quad (9)$$

Thus, we can obtain the element $f(u_i|\boldsymbol{z})$ $(i = 1, 2, \cdots, k)$ of $M$ from equation (9), which is determined uniquely according to the order of disjoint solution $\lambda_1 < \cdots < \lambda_k$ of equation (8). Inversely, since the order of the elements of $M$ is identifiable from condition (ii), the order of $\lambda_1, \cdots, \lambda_k$ is identifiable. Thus, the conditional distribution of $U$ given $\boldsymbol{z}$ is estimable through the observation of $\boldsymbol{S} \cup \boldsymbol{T} \cup \boldsymbol{W} \cup \boldsymbol{Z}$. Then, since

$$f(u, \boldsymbol{z}, \boldsymbol{w}) = f(\boldsymbol{w}|u, \boldsymbol{z}) f(u|\boldsymbol{z}) f(\boldsymbol{z}),$$

$f(u, \boldsymbol{z}, \boldsymbol{w})$ is estimable through the observation of $\boldsymbol{S} \cup \boldsymbol{T} \cup \boldsymbol{W} \cup \boldsymbol{Z}$. \hfill Q.E.D.

Based on Theorem 1, the following corollary can be derived immediately.

**COROLLARY 1**
Suppose that one element of $\{X, Y\}$ is an unobserved variable $U$ and the other element is included in a set $\boldsymbol{Z}\cup\boldsymbol{W}$ of observed variables. Let $\boldsymbol{C}$ be a subset of $\boldsymbol{Z}\cup\boldsymbol{W}\backslash\{X,Y\}$ that satisfies the identifiability criteria for the causal effect $f(y|\text{set}(X=x))$. If a set $\boldsymbol{Z}\cup\boldsymbol{W}\cup\boldsymbol{S}\cup\boldsymbol{T}$ of observed variables satisfies conditions (i)-(iii) in Theorem 1, then the causal effect $f(y|\text{set}(X=x))$ is identifiable. □

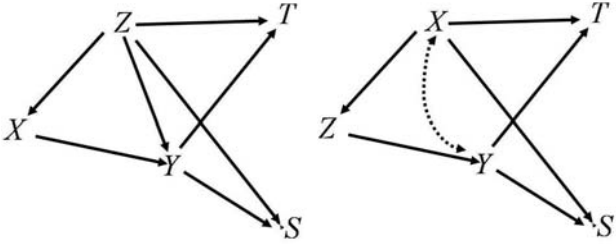

Fig. 1 : Causal diagram (1)   Fig. 2 : Causal diagram (2)

We use two examples to illustrate Corollary 1. First, consider the causal diagram shown in Fig. 1. Setting $\boldsymbol{W}$ in Corollary 1 to $X$ in Fig. 1, we can recognize that $\{Z,Y\}$ d-separates $S$ from $T$ and $X$ from $\{S,T\}$. In addition, $\boldsymbol{C}=\{Z\}$ satisfies the back door criterion relative to $(X,Y)$. Then, based on the proof of Theorem 1, the distribution of $X,Y$ and $Z$ can be constructed according to the distribution of $X,Z,S$ and $T$. Thus, if conditions (ii) and (iii) hold true, then the causal effect $f(y|\text{set}(X=x))$ is identifiable through the observation of $X,S,T$ and $Z$. The closed form expression in the case where $Y$ is a dichotomous variable is provided in Kuroki (2007).

Next, consider the causal diagram shown in Fig. 2, where the back door criterion cannot be applied to identify the causal effect of $f(y|\text{set}(X=x))$, because there is a bi-directed arrow in Fig. 2 which indicates that there exist some unmeasured confounders between $X$ and $Y$. Letting $U$, $\boldsymbol{W}$ and $\boldsymbol{Z}$ in Corollary 1 be $Y$, $\phi$ and $X$ in Fig. 2, we can recognize that $\{X,Y\}$ d-separates $S$ from $T$ and $Z$ from $\{S,T\}$. In addition, $\boldsymbol{C}=\{Z\}$ satisfies the front door criterion relative to $(X,Y)$. Then, based on the proof of Theorem 1, the distribution of $X,Y$ and $Z$ can be constructed according to the distribution of $X,Z,S$ and $T$. Thus, if conditions (ii) and (iii) hold true, then the causal effect $f(y|\text{set}(X=x))$ is identifiable through the observation of $X,S,T$ and $Z$. This example shows that our result can also be applied to situations where there is no variable that satisfies the back door criterion.

When identifying the causal effect using Theorem 1, it is required that $f(u_1|\boldsymbol{z}) < f(u_2|\boldsymbol{z}) < \cdots < f(u_k|\boldsymbol{z})$ holds true for any $\boldsymbol{z}$. If such an order information is not available, it is impossible to judge whether the causal effect is identifiable or not from Theorem 1. But we can evaluate the bounds of the causal effect. Consider the causal diagram shown in Fig. 1 as an example. By noting that $f(y|x,z) = f(x|y,z)f(y|z)/f(x|z)$ holds

true, letting the diagonal elements of $M$ be $m_1,\cdots,m_k$ determined according to the order $\lambda_1 < \cdots < \lambda_k$, the bounds of the causal effect $f(y|\text{set}(X=x))$ are

$$\sum_z \frac{\min_i\{\lambda_i m_i\}}{f(x|z)} f(z) \leq f(y|\text{set}(X=x))$$

$$f(y|\text{set}(X=x)) \leq \sum_z \frac{\max_i\{\lambda_i m_i\}}{f(x|z)} f(z).$$

## 4 BOUNDS ON CAUSAL EFFECT

### 4.1 POTENTIAL OUTCOME APPROACH

In this section, we consider the case where there exist unmeasured variables between an unobserved outcome variable and its proxy variables. Under such a situation, it is impossible to identify the causal effect, but we can derive the bounds on the causal effect by using the potential outcome approach. For simplicity, we only consider the case of an unobserved dichotomous outcome variable, though our result can apply to multi-categorical case directly.

Let $X$ and $Y$ be a dichotomous exposure variable ($x \in \{x_0, x_1\}$) and a dichotomous outcome variable ($y \in \{y_0, y_1\}$). Then, the $i$th of the $N$ subjects has both an outcome $Y_{x_1}(i)$ that have resulted if he was exposed to $x_1$, and an outcome $Y_{x_0}(i)$ that have resulted if he was exposed to $x_0$. When the $N$ subjects in the study are considered as a random sample from the target population, since $Y_{x_1}(i)$ and $Y_{x_0}(i)$ can be referred to as the values of random variables $Y_{x_1}$ and $Y_{x_0}$ respectively, the causal effect can be defined as the probability $P(Y_x = y) \triangleq f(y_x)$ of the potential outcome ($x \in \{x_0, x_1\}$). The potential outcome $Y_x$ is observed only if the subject receives exposure $x$ ($x \in \{x_1, x_0\}$). Thus, when randomized experiment is conducted and compliance is perfect, the causal effect of $X$ on $Y$ is

$$f(y_x) \triangleq f(y|\text{set}(X=x)) = f(y|x), \qquad (10)$$

by using the consistency condition (Pearl, 2000)

$$X = x \Rightarrow Y_x = Y.$$

On the other hand, when randomized experiment is difficult to conduct and only observational data is available, we can still estimate the causal effect according to the strongly-ignorable-treatment-assignment (SITA) condition (Rosenbaum and Rubin, 1983). That is, for the exposure variable $X$, if there exists such a set $\boldsymbol{Z}$ of covariates that $X$ is conditionally independent of $(Y_{x_1}, Y_{x_0})$ given $\boldsymbol{Z}$, denoted as $X \perp\!\!\!\perp (Y_{x_1}, Y_{x_0}) | \boldsymbol{Z}$, we shall say treatment assignment is strongly ignorable given $\boldsymbol{Z}$, or $\boldsymbol{Z}$ satisfies the SITA condition. Thus, $f(y|\text{set}(X=x))$ is estimable by using $Z$ and is given as equation (3).

## 4.2 FORMULATION

In order to describe our problem, we consider the simple causal diagram shown in Fig. 3, where $X$, $S$ and $T$ are observed dichotomous variables ($x \in \{x_0, x_1\}, s \in \{s_0, s_1\}, t \in \{t_0, t_1\}$). In addition, $Y$ is an unobserved dichotomous variable ($y \in \{y_0, y_1\}$). Furthermore, there is no confounder between $X$ and $Y$, but there exist unmeasured variables between $Y$, $S$ and $T$.

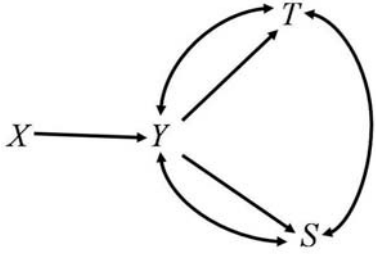

Fig. 3: Causal Diagram (3)

Then, the potential outcomes corresponding to this figure can be introduced as follows:

(i) Potential outcome $R_t$ in the context of $Y$ as an exposure and $T$ as an outcome:

$r_{t_0}:(T_{y_0}, T_{y_1}) = (t_0, t_0), r_{t_1}:(T_{y_0}, T_{y_1}) = (t_0, t_1),$

$r_{t_2}:(T_{y_0}, T_{y_1}) = (t_1, t_0), r_{t_3}:(T_{y_0}, T_{y_1}) = (t_1, t_1),$

(ii) Potential outcome $R_s$ in the context of $Y$ as an exposure and $S$ as an outcome:

$r_{s_0}:(S_{y_0}, S_{y_1}) = (s_0, s_0), r_{s_1}:(S_{y_0}, S_{y_1}) = (s_0, s_1),$

$r_{s_2}:(S_{y_0}, S_{y_1}) = (s_1, s_0), r_{s_3}:(S_{y_0}, S_{y_1}) = (s_1, s_1),$

(iii) Potential outcome $R_y$ in the context of $X$ as an exposure and $Y$ as an outcome:

$r_{y_0}:(Y_{x_0}, Y_{x_1}) = (y_0, y_0), r_{y_1}:(Y_{x_0}, Y_{x_1}) = (y_0, y_1),$

$r_{y_2}:(Y_{x_0}, Y_{x_1}) = (y_1, y_0), r_{y_3}:(Y_{x_0}, Y_{x_1}) = (y_1, y_1).$

Letting $q_{ijk} = P(r_{t_i}, r_{s_j}, r_{y_k})$ be counterfactual probabilities ($i, j, k = 0, 1, 2, 3$), these parameters are constrained by the probabilistic equality

$$\sum_{i=0}^{3}\sum_{j=0}^{3}\sum_{k=0}^{3} q_{ijk} = 1 \text{ and } 0 \leq q_{ijk} \leq 1. \quad (11)$$

Let $p_{ij \cdot k}$ be the observed conditional probabilities of $(T, S) = (t_i, s_j)$ given $X = x_k$, that is, $p_{ij \cdot k} = P(t_i, s_j | x_k)$. These observed conditional probabilities impose the constraints by applying both the consistency condition and $X \perp\!\!\!\perp (S_{y_0}, S_{y_1}, T_{y_0}, T_{y_1}, Y_{x_0}, Y_{x_1})$ to the counterfactual probabilities:

$p_{00 \cdot 1} = \sum_{i=0,1}\sum_{j=0,1}\sum_{k=0,2} q_{ijk} + \sum_{i=0,2}\sum_{j=0,2}\sum_{k=1,3} q_{ijk},$

$p_{01 \cdot 1} = \sum_{i=0,1}\sum_{j=2,3}\sum_{k=0,2} q_{ijk} + \sum_{i=0,2}\sum_{j=1,3}\sum_{k=1,3} q_{ijk}$

$p_{10 \cdot 1} = \sum_{i=2,3}\sum_{j=0,1}\sum_{k=0,2} q_{ijk} + \sum_{i=1,3}\sum_{j=0,2}\sum_{k=1,3} q_{ijk},$

$p_{11 \cdot 1} = \sum_{i=2,3}\sum_{j=2,3}\sum_{k=0,2} q_{ijk} + \sum_{i=1,3}\sum_{j=1,3}\sum_{k=1,3} q_{ijk}$

$p_{00 \cdot 0} = \sum_{i=0,1}\sum_{j=0,1}\sum_{k=0,1} q_{ijk} + \sum_{i=0,2}\sum_{j=0,2}\sum_{k=2,3} q_{ijk},$

$p_{01 \cdot 0} = \sum_{i=0,1}\sum_{j=2,3}\sum_{k=0,1} q_{ijk} + \sum_{i=0,2}\sum_{j=1,3}\sum_{k=2,3} q_{ijk}$

$p_{10 \cdot 0} = \sum_{i=2,3}\sum_{j=0,1}\sum_{k=0,1} q_{ijk} + \sum_{i=1,3}\sum_{j=0,2}\sum_{k=2,3} q_{ijk},$

$p_{11 \cdot 0} = \sum_{i=2,3}\sum_{j=2,3}\sum_{k=0,1} q_{ijk} + \sum_{i=1,3}\sum_{j=1,3}\sum_{k=2,3} q_{ijk}.$

(12)

Then, the quantities we wish to bound are:

$$f(y_1|\text{set}(X = x_1)) = \sum_{i=0}^{3}\sum_{j=0}^{3}\sum_{k=1,3} q_{ijk}, \quad (13)$$

$$f(y_1|\text{set}(X = x_0)) = \sum_{i=0}^{3}\sum_{j=0}^{3}\sum_{k=2,3} q_{ijk}. \quad (14)$$

Optimizing the functions (13) and (14), subject to equality constraints (11) and (12), defines a linear programming (LP) problem that lends itself to closed-form solution. Balke (1995) describes a computer program that takes symbolic description of LP problems and returns symbolic expressions for the desired bounds. The problem works by systematically enumerating the vertices of the constraint polygon of the dual problem. The bounds reported in this paper were produced by using Balke's program, and will be stated here without proofs; their correctness can be verified by manually enumerating the vertices as described in Balke (1995). These bounds are guaranteed to be sharp because the optimization is global.

Given the observed conditional probabilities, the constraints (11) and (12) induce the bounds [0, 1], which indicates that the causal knowledge available from Fig. 3 can not provide useful evaluation of the causal effect. However, if we assume the monotonic assumption, which leads to

$$q_{2jk} = q_{i2k} = q_{ij2} = 0 \quad i, j, k = 0, 1, 2, 3.$$

Then, we can obtain the tightest bounds on the causal effects:

$$0 \leq f(y_1|\text{set}(X = x_0)) \leq$$
$$\min \left\{ \begin{array}{l} p_{01 \cdot 0} + p_{10 \cdot 0} + p_{11 \cdot 0} + p_{00 \cdot 1} \\ p_{01 \cdot 0} + p_{11 \cdot 0} + p_{10 \cdot 1} + p_{00 \cdot 1} \\ p_{10 \cdot 0} + p_{11 \cdot 0} + p_{01 \cdot 1} + p_{00 \cdot 1} \\ p_{11 \cdot 0} + p_{00 \cdot 1} + p_{10 \cdot 1} + p_{01 \cdot 1} \end{array} \right\}, \quad (15)$$

$$\max \left\{ \begin{array}{c} p_{00 \cdot 0} - p_{00 \cdot 1} \\ p_{11 \cdot 1} - p_{11 \cdot 0} \\ p_{00 \cdot 0} + p_{10 \cdot 0} - p_{00 \cdot 1} - p_{10 \cdot 1} \\ p_{00 \cdot 0} + p_{01 \cdot 0} - p_{00 \cdot 1} - p_{01 \cdot 1} \end{array} \right\}$$
$$\leq f(y_1|\text{set}(X = x_1)) \leq 1. \qquad (16)$$

It is noted that these formulas require two proxy variables $S$ and $T$. If only one proxy variable is available, the tightest bounds on the causal effects become $[0, 1]$, which shows that one proxy variable provides no useful information for evaluating the causal effect.

Finally, we consider a more complicated situation that there are confounders between $X$ and $Y$. In the case, if we can observe a set $\boldsymbol{Z}$ of covariates that satisfy the SITA condition, and $X \perp\!\!\!\perp (S_{y_0}, S_{y_1}, T_{y_0}, T_{y_1}, Y_{x_0}, Y_{x_1})|\boldsymbol{Z}$ holds true, by the same procedure as the above, the bounds on the causal effects can be evaluated as

$$0 \leq f(y_1|\text{set}(X = x_0)) \leq$$
$$\sum_z \min \left\{ \begin{array}{c} p_{01 \cdot 0z} + p_{10 \cdot 0z} + p_{11 \cdot 0z} + p_{00 \cdot 1z} \\ p_{01 \cdot 0z} + p_{11 \cdot 0z} + p_{10 \cdot 1z} + p_{00 \cdot 1z} \\ p_{10 \cdot 0z} + p_{11 \cdot 0z} + p_{01 \cdot 1z} + p_{00 \cdot 1z} \\ p_{11 \cdot 0z} + p_{00 \cdot 1z} + p_{10 \cdot 1z} + p_{01 \cdot 1z} \end{array} \right\} P(z),$$

$$\sum_z \max \left\{ \begin{array}{c} p_{00 \cdot 0z} - p_{00 \cdot 1z} \\ p_{11 \cdot 1z} - p_{11 \cdot 0z} \\ p_{00 \cdot 0z} + p_{10 \cdot 0z} - p_{00 \cdot 1z} - p_{10 \cdot 1z} \\ p_{00 \cdot 0z} + p_{01 \cdot 0z} - p_{00 \cdot 1z} - p_{01 \cdot 1z} \end{array} \right\} P(z)$$
$$\leq f(y_1|\text{set}(X = x_1)) \leq 1.$$

With the similar procedure above, we can also derive the bounds on the causal effect when there exist unmeasured variables between an unobserved exposure variable and its proxy variables.

## 5 EXAMPLE

We illustrate our results using the data from a political action study reported by Hagenaars (1993). He analyzed the data in order to evaluate the causal effect of education on political involvement. The variables of interest are the following:

$X$: education (1: some college; 2: less than college),

$S$: ideological level (1: ideologues; 2: nonideologues),

$T$: repression potential (1: low; 2: high),

$Y$: political involvement (1: high; 2: low),

Here, we concentrate our discussion on evaluating the causal effect of $X$ on $Y$, where $Y$ is unobserved. In order to help readers understand our results, we consider a submodel in Hagenaars (1993), which is shown in Fig. 3.

First, we consider the situation where there is no bidirected arrow in Fig. 3. Since $Y$ d-separates any two vertices in $\{S, T, X\}$, and there is no unmeasured variables between $X$ and $Y$, Theorem 1 can be used to achieve our aim. Because the real data of this model is not available, we generate hypothetical data according to Fig. 3, which is shown in Table 1.

Table 1. Hypothetical Data of the Example

|       |       | $t_1$  | $t_0$  |
|-------|-------|--------|--------|
| $x_1$ | $s_1$ | 0.0648 | 0.0432 |
|       | $s_0$ | 0.1392 | 0.0528 |
| $x_0$ | $s_1$ | 0.1092 | 0.2478 |
|       | $s_0$ | 0.1568 | 0.1862 |

In this example, we suppose that $f(u_1) < f(u_2)$. Then, letting

$$P = \left( \begin{array}{cc} 1.000 & f(t_1) \\ f(s_1) & f(s_1, t_1) \end{array} \right) = \left( \begin{array}{cc} 1.000 & 0.47 \\ 0.465 & 0.174 \end{array} \right)$$

$$Q = \left( \begin{array}{cc} f(x_1) & f(t_1, x_1) \\ f(s_1, x_1) & f(s_1, t_1, x_1) \end{array} \right) = \left( \begin{array}{cc} 0.3 & 0.204 \\ 0.108 & 0.0648 \end{array} \right),$$

the eigenvalues of $P^{-1}Q$ and $P'^{-1}Q'$ are 0.533 and 0.109, and the corresponding eigenmatrices are

$$A_1 = \left( \begin{array}{cc} 0.196 & -0.625 \\ -0.981 & 0.781 \end{array} \right), A_2 = \left( \begin{array}{cc} 0.514 & -0.287 \\ -0.857 & 0.958 \end{array} \right),$$

for $P^{-1}Q$ and $P'^{-1}Q'$ respectively. Thus, letting $E_i = \text{diag}(a_{1(i)}, a_{2(i)})$ and

$$P_1 = \left( \begin{array}{cc} 1.000 & f(s_1|y_1) \\ 1.000 & f(s_1|y_2) \end{array} \right), P_2 = \left( \begin{array}{cc} 1.000 & f(t_1|y_1) \\ 1.000 & f(t_1|y_2) \end{array} \right),$$

since we can provide $E_1 = \text{diag}(-0.588, -0.469)$ and $E_2 = \text{diag}(0.257, 0.287)$, noting that $P_1 = E_1 A_1^{-1}$ and $P_2 = E_2 A_2^{-1}$, the followings can be derived:

$$P_1 = \left( \begin{array}{cc} 1.000 & 0.800 \\ 1.000 & 0.200 \end{array} \right), P_2 = \left( \begin{array}{cc} 1.000 & 0.300 \\ 1.000 & 0.600 \end{array} \right).$$

Thus, $M = \text{diag}(f(u_1), f(u_2)) = P_2'^{-1} P P_1^{-1} = \text{diag}(0.45, 0.55)$ and the causal effects of $X$ on $Y$ are $f(y_1|\text{set}(X = x_1)) = 0.533 \times 0.45/0.3 = 0.8$ and $f(y_1|\text{set}(X = x_0)) = 1.000 - (1.000 - 0.109) \times 0.55/0.7 = 0.3$, respectively.

Second, we consider the situation where there are bidirected arrows shown in Fig. 3. Under this situation, we can not evaluate the causal effects of $X$ on $Y$ by Theorem 1. However, under the monotonic assumption, we can provide the tightest bounds on the causal effects as

$$0.000 < P(y_1|\text{set}(X = x_0))$$
$$< \min\{0.567, 0.549, 0.384, 0.478\} = 0.384$$

$$\max\{0.205, -0.044, 0.151, 0.338\} = 0.338$$
$$< P(y_1|\text{set}(X = x_0)) < 1.000,$$

based on equations (15) and (16).

## 6 DISCUSSION

This paper derived the graphical identifiability criteria for causal effects based on causal modeling in observational studies with an unobserved multi-categorical exposure/outcome variable. In addition, when unmeasured variables exist between an unobserved outcome variable and its proxy variables, we provided the tightest bounds on causal effects by using Balke's LP program method. The results of this paper enable us to evaluate causal effects when it is difficult to observe an exposure/outcome variable.

**ACKNOWLEGDEMENT**

This research was partly supported by the Kurata Foundation, the Mazda Foundation and the Ministry of Education, Culture, Sports, Science and Technology of Japan.